\documentclass{article}
\usepackage[T1]{fontenc} 
\usepackage[utf8]{inputenc} 
\usepackage{ismir,amsmath,cite,url}
\usepackage{graphicx}
\usepackage{color}

\usepackage{amsfonts}
\usepackage{multirow, amsmath, booktabs}
\usepackage{multicol}
\usepackage[dvips]{epsfig}

\usepackage{makecell}
\usepackage{float}

\title{Investigating U-Nets with various Intermediate Blocks for Spectrogram-based Singing Voice Separation}

\multauthor
{Woosung Choi$^1$ \hspace{1cm} Minseok Kim$^1$ \hspace{1cm} Jaehwa Chung$^2$} { \bfseries{Daewon Lee$^3$ \hspace{1cm} Soonyoung Jung$^1$}\\
  $^1$ Department of Computer Science and Engineering, Korea University, Republic of Korea\\
$^2$ Department of Computer Science, Korea National Open University, Republic of Korea\\
$^3$  Department of Computer Engineering, Seokyeong University, Republic of Korea\\
{\tt\small jsy@korea.ac.kr}
}
\def\authorname{Woosung Choi, Minseok Kim, Jaehwa Chung, Daewon Lee, Soonyoung Jung}

\begin{document}

\maketitle
\begin{abstract}

Singing Voice Separation (SVS) tries to separate singing voice from a given mixed musical signal. 
Recently, many U-Net-based models have been proposed for the SVS task, but there were no existing works that evaluate and compare various types of intermediate blocks that can be used in the U-Net architecture.
In this paper, we introduce a variety of intermediate spectrogram transformation blocks. We implement U-nets based on these blocks and train them on complex-valued spectrograms to consider both magnitude and phase. These networks are then compared on the SDR metric. When using a particular block composed of convolutional and fully-connected layers, it achieves state-of-the-art SDR on the MUSDB singing voice separation task by a large margin of 0.9 dB. Our code and models are available online. \footnote{https://github.com/ws-choi/ISMIR2020\_U\_Nets\_SVS} 

\end{abstract}
\section{Introduction}
\label{sec:intro}

Singing Voice Separation (SVS), a special case of Music Source Separation (MSS), aims at separating singing voice from a given mixed musical signal.
Recently, many machine learning-based methods have been proposed for SVS and MSS tasks.
They can be categorized into two groups: waveform-to-waveform models and spectrogram-based models.
While the former tries to generate the vocal waveforms directly, the latter estimates spectrograms (usually magnitude) of vocal waveforms.

Typical spectrogram-based models apply Short-Time Fourier Transform (STFT) on a mixture waveform to obtain the input spectrograms.
Then, they estimate the vocal spectrograms based on these inputs and finally restore the vocal waveform with inverse STFT (iSTFT).
A variety of spectrogram-based models have been proposed in the music information retrieval community and the machine learning community.
For example, \cite{unet} employed the U-Net \cite{medicalunet} architecture, an encoder-decoder structure with symmetric skip connections.
These symmetric skip connections allow models to recover fine-grained details of the target object during decoding effectively.
Several works \cite{shn,sashn,mmdensenet,mmdenselstm} also used similar architectures.

They have revealed that U-Net-like architectures can provide promising performance for SVS and MSS.
Existing works have proposed various types of neural networks for intermediate blocks. 
While some models \cite{unet,shn} used simple Convolutional Neural Networks (CNNs) for intermediate blocks, other advanced models tried more complex intermediate blocks.
For instance, MMDenseLSTM \cite{mmdenselstm} used densely connected CNNs followed by Long Short-Term Memory (LSTM) networks to efficiently model long-term structures, where  LSTM is a variant of Recurrent Neural Networks (RNNs).
However, a thorough search of the relevant literature indicated that there were no existing works that evaluate and directly compare these different types of blocks.

In this paper, we conduct a comparative study of U-Nets on various intermediate blocks.
We designed several types of blocks based on different design strategies, which we present in section \ref{sec:blocks}. 
For each type of block, we implemented at least one SVS model, which are all based on an identical U-Net framework for fair comparisons. 
In section \ref{sec:exp}, we summarize the experimental results and discuss the effect of each design choice. 
We validate hypotheses such as that inserting time-distributed operations (see \S \ref{sec:tib}) into intermediate blocks can significantly improve performance, which led to state-of-the-art (SOTA) performance on the MUSDB \cite{musdb18} SVS task.

Finally, our U-Net framework directly estimates the target complex-valued spectrogram (viewing real and imaginary as separate channels), when many existing models estimate the target magnitude without phase. In general, considering phase information improves the separation quality, as discussed in \cite{phasen,phase2}. Several phase-aware methods have been proposed for speech enhancement, such as phase reconstruction methods \cite{phasen,phase2}, or using raw complex-valued STFT outputs \cite{complex,cac1}. In section \ref{sec:exp}, we show that the latter method is an efficient way to improve magnitude-only models, only needing a few minor adjustments.

\section{U-Net-based SVS Framework}
\label{sec:framework}

In this section, we describe a U-Net-based SVS framework, which is shared by several models in \S \ref{sec:exp}.
We first introduce the `Complex as Channel framework' (CaC), a spectrogram-based SVS framework, and then define our U-Net architecture for spectrogram estimation in CaC.

\subsection{Complex as Channel Framework}
\label{sec:cac}

\begin{figure}[htbp]
\centering
\includegraphics[width=\columnwidth]{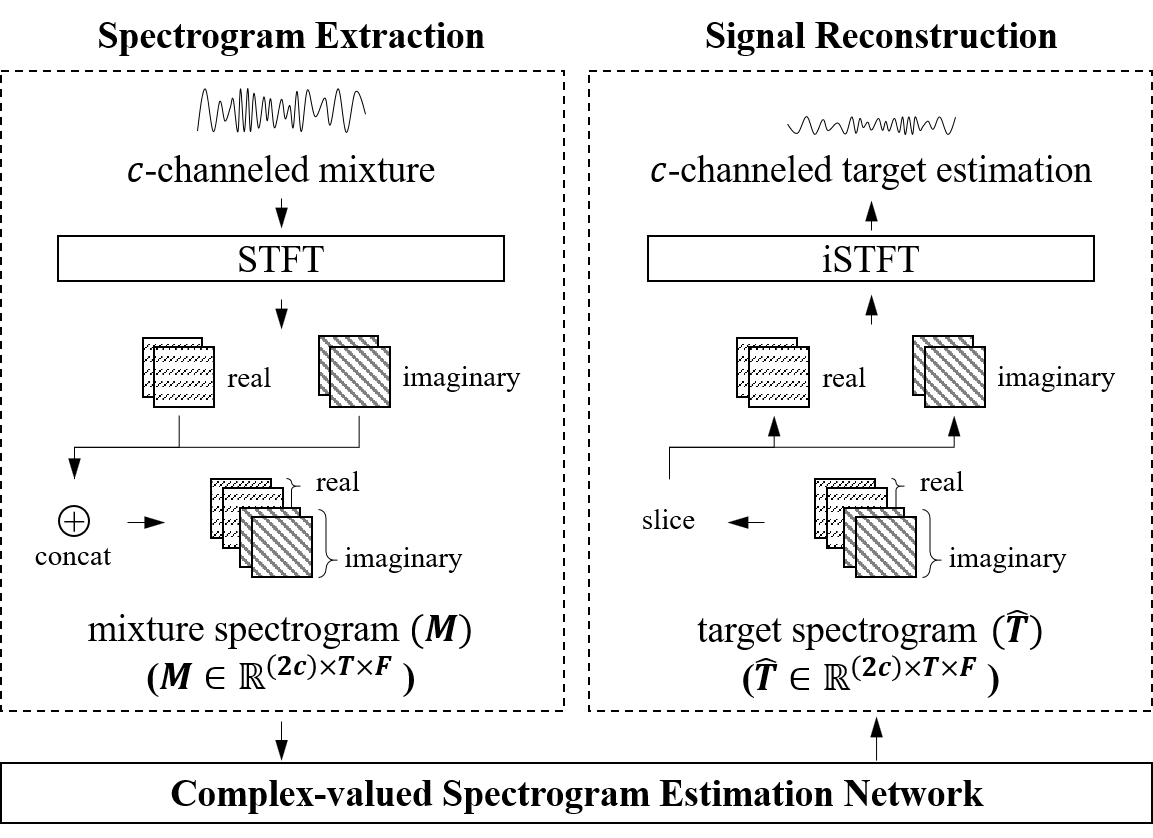}
\caption{The Complex as Channel Framework}
\label{fig:cac}
\end{figure}

CaC is a singing voice separation framework based on complex-valued spectrogram estimation.
It takes a $c$-channeled mixture signal, and outputs $c$-channeled singing voice signal.
As shown in Figure \ref{fig:cac}, CaC consists of three parts as follows: 

\begin{enumerate}
    \item The \textit{spectrogram extraction layer} extracts a mixture spectrogram by applying STFT to the $c$-channeled input signal. The output of STFT is a complex-valued spectrogram with $c$-channels. Considering the imaginary and real parts as separate real-valued channels, we view the mixture spectrogram $M_{complex} \in \mathbb{C} ^{c \times T \times F}$ as a $(2c)$-channeled real-valued spectrogram $M \in \mathbb{R} ^{ \left(2c\right) \times T \times F}$, where $T$ denotes the number of frames and $F$ denotes the number of the frequency bins in the spectrogram. 
    
    \item The \textit{complex-valued spectrogram estimation network} is a  neural network that takes the spectrogram $\mathbf{M}$ of a mixture signal as input and estimates the target spectrogram $\hat{T} \in \mathbb{R} ^{\; \left(2c\right) \times T \times F}$, which is used for reconstructing the vocal signal later. 
    
    \item The \textit{signal reconstruction layer} reshapes the estimated spectrogram $\hat{T}$ into the complex-valued spectrogram  $\hat{T}_{complex} \in \mathbb{C} ^{\; c \times T \times F}$, as shown in Figure \ref{fig:cac}. It then restores the estimated singing voice signal via inverse-STFT on $\hat{T}_{complex}$.

\end{enumerate}

For a given mixture spectrogram $\mathbf{M}$, we train the complex-valued spectrogram estimation network in a supervised fashion to minimize the mean square error between the output $\hat{T}$ and the ground-truth spectrogram $\mathbf{T}$ of the singing voice signal.

It should be noted that the shape of $\mathbf{M}$ and $\hat{T}$ is $\left( 2c \right) \times T \times F$, considering real and imaginary parts of a spectrogram as separate real-valued channels.
This approach allows CaC to fully utilize the information in complex-valued spectrograms for both the input and the output.
Meanwhile, current SOTA models (e.g., SA-SHN \cite{sashn} and DGRU-DGConv \cite{dilatedlstm}) decompose a complex-valued spectrogram into magnitude and phase, and only use the magnitude for the input of their networks. 
Although SA-SHN and DGRU-DGConv yielded impressive results by introducing novel attention method \cite{sashn} and by adopting dilated 1-D convolutions \cite{dilatedlstm} with Gated Recurrent Units (GRU) \cite{gru} respectively, they do not consider phase information.
In \S \ref{sec:cvsm}, we compare the Source-to-Distirtion (SDR) \cite{bss} performance of models based on the CaC framework and that of models based on the Magnitude-only framework.

\subsection{U-Net Architecture for Spectrogram Estimation}
\label{sec:unet}

For spectrogram estimation in CaC, we use a U-Net-based architecture. 
It consists of an encoder and a decoder: the encoder transforms $\mathbf{M}$ into a downsized spectrogram-like representation, and the decoder takes it and returns the estimated target spectrogram $\hat{T}$. Before we describe them in detail, we introduce two types of main components in the architecture as follows.

\begin{itemize}
    \item An \textit{intermediate block} transforms an input spectrogram-like tensor into an equally-sized tensor (possibly with a different number of channels).
    \item A \textit{down/up sampling layer} halves/doubles the scale of an input tensor either in the time, frequency, or Time-Frequency domain.
\end{itemize}

\begin{figure}[htbp]
\centering
\includegraphics[width=\columnwidth]{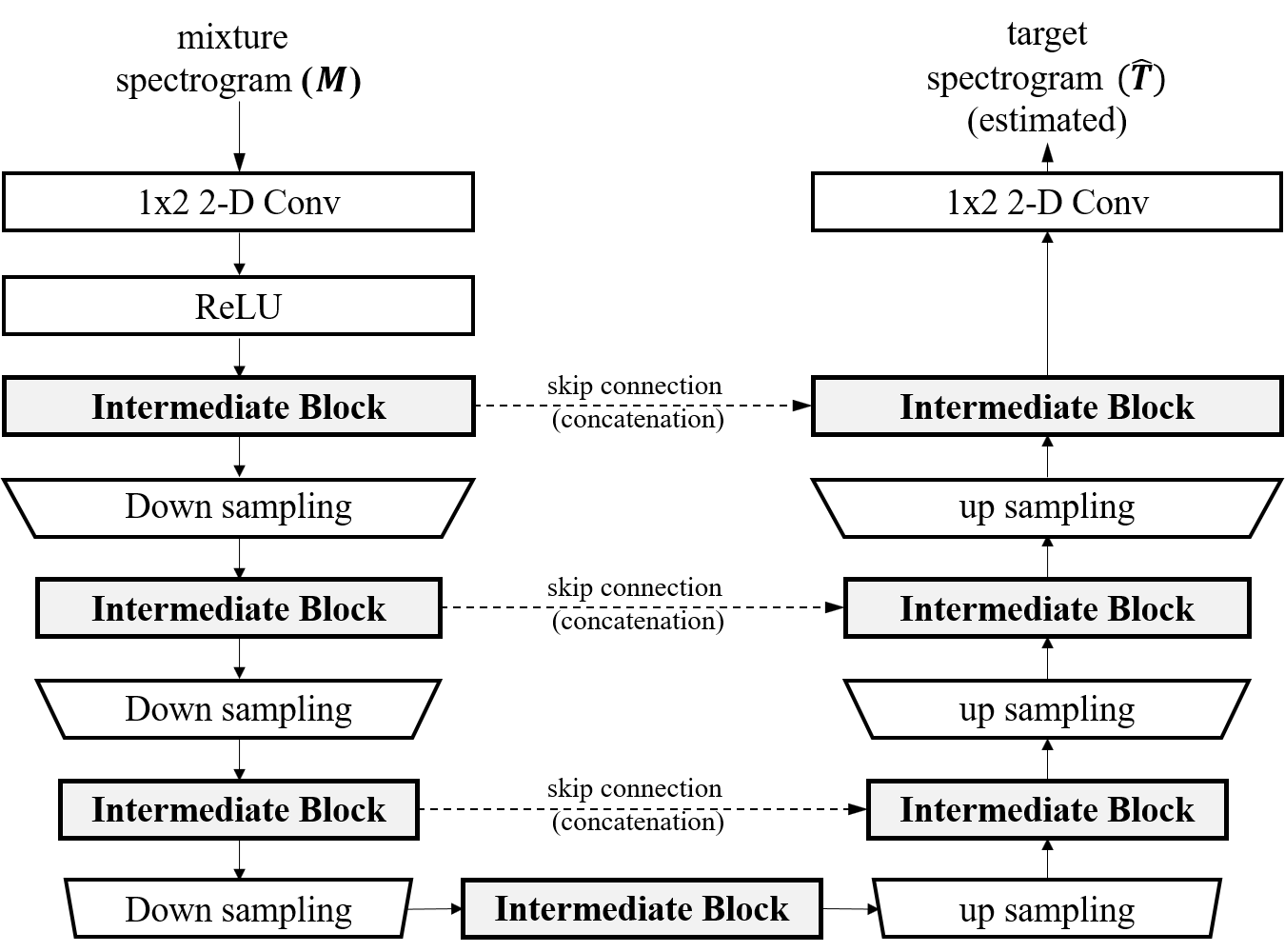}
\caption{U-Net Architecture for Spectrogram Estimation}
\label{fig:unet}
\end{figure}

As shown in Figure \ref{fig:unet}, the number of down-sampling layers and the number of up-sampling layers are the same.
Also, it uses the same number of intermediate blocks in the encoding and the decoding phase. It has an additional block in between its encoder and decoder.
Thus, the total number of blocks should be an odd integer.
It has skip connections that concatenate output feature maps of the same scale between the encoder and the decoder.

Besides basic components, our architecture has two additional convolution layers, as illustrated in Figure \ref{fig:unet}.
We use them to increase or restore the number of channels.
Before describing them, let us introduce some notations.
We denote the input of the $l$-th intermediate block by $X^{(l-1)}$, and the output by $X^{(l)}$.
The size of $X^{(l-1)}$ is denoted by $c_{in}^{(l)} \times T^{(l)} \times F^{(l)}$, where $c_{in}^{(l)}$ represents the number of channels and and $T^{(l)} \times F^{(l)}$ represents the size of the spectrogram-like tensor.
Also, we denote the size of $X^{(l)}$ by $c_{out}^{(l)} \times T^{(l)} \times F^{(l)}$, where $c_{out}^{(l)}$  is the number of channels. 
Using these notations, we denote the input of the first  block by $X^{(0)}$, and its size by $c_{in}^{(1)} \times T^{(1)} \times F^{(1)}$. 
To increase the number of channels, it applies a $1\times 2$ convolution with $c_{in}^{(1)}$ output channels followed by ReLU \cite{relu} activation to the given input $\mathbf{M}$.
To adjust the number of channels, it also applies a final $1\times 2$ convolution with $(2c)$ output channels to the output of the final block. 
Note that the last layer is not followed by any activation function since target TF bins can be negative.
We empirically set the parameter  $c_{in}^{(1)}$ to be 24 in our experiments. Models with smaller $c_{in}^{(1)}$ (e.g., 12) are trained faster, but usually perform inferior than models with larger size of $c_{in}^{(1)}$.

We can implement various SVS models based on this architecture in the CaC framework because multiple options are available for intermediate blocks.
In section \ref{sec:blocks}, we present several neural networks which can be used as intermediate blocks in this paper.

\section{Intermediate Blocks}

\label{sec:blocks}

We present several types of intermediate blocks based on different design strategies. We first present time-distributed blocks and then present time-frequency blocks.

\subsection{Time-Distributed Blocks}
\label{sec:tib}
Some existing models use CNNs (e.g., \cite{ticbase}) for intermediate blocks to extract timbre features of the target source.
However, the authors of \cite{phasen} reported that conventional CNN kernels are limited for this task.
They found that long-range correlations exist along the frequency axis in the spectrogram of voice signals, which Fully-connected Neural Networks (FCNs) can efficiently capture.
They proposed a model named Phasen for speech enhancement, which uses the Frequency Transformation Block (FTB) that has a single-layered FCN without bias. This FCN is applied to each frame of the internal representation in a \textit{ time-distributed} manner. 

Inspired by TFB, we introduce \textit{time-distributed} blocks, which are applied to a single frame of a spectrogram-like feature map. These blocks try to extract time-independent features that help singing voice separation without using inter-frame operations.
We first introduce an FCN-based block and then propose an alternative time-distributed block based on 1-D CNNs. 

\subsubsection{Time-Distributed Fully-connected networks}
\label{sec:tdf}

\begin{figure}[htbp]
    \centering
    \includegraphics[width=\columnwidth]{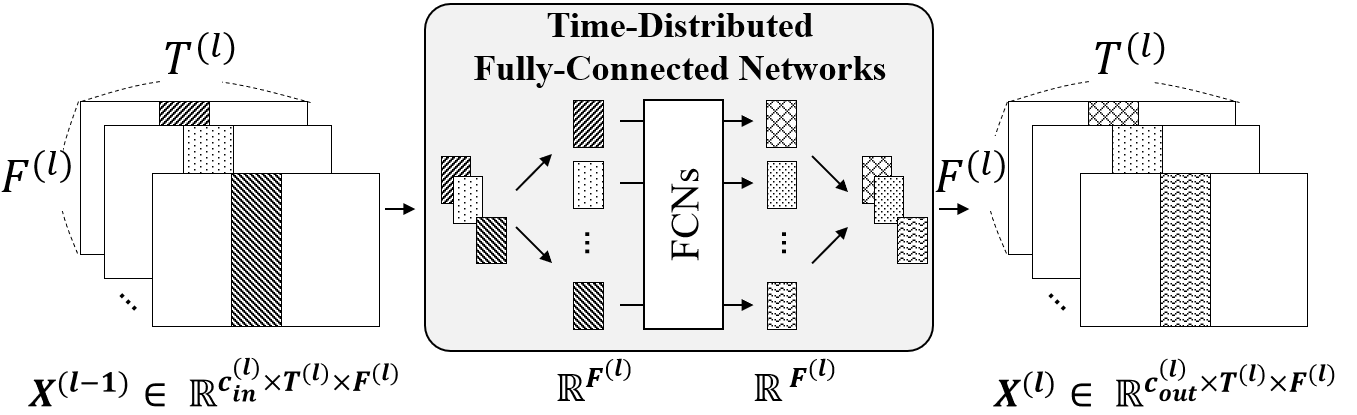}
    \caption{Time-Distributed Fully-connected networks}
    \label{fig:tdf}
\end{figure}

We present an FCN-based intermediate block, called Time-Distributed Fully-connected network (TDF). 
As illustrated in Figure \ref{fig:tdf}, a TDF block is applied to each channel of each frame separately and identically. 

Suppose that the $l$-th intermediate block in our U-Net structure takes input $X^{(l-1)}$ into an output $X^{(l)}$.
As shown in Figure \ref{fig:tdf}, a  fully-connected network is applied separately and identically to each frame (i.e., $X^{(l-1)}[i,j,:]$) in order to transform an input tensor in a time-distributed fashion. 
While an FTB of Phasen \cite{phasen} is single-layered, a TDF block can be either single- or multi-layered. Each layer is defined as consecutive operations: a fully-connected layer, Batch Norm (BN) \cite{bn}, and ReLU \cite{relu}. If it is multi-layered, then each internal layer maps an input to the hidden feature space, and its final layer maps the internal vector to $\mathbb{R}^{F^{(l)}}$.
The number of hidden units is $\lfloor F^{(l)}/bn \rfloor$, where we denote the bottleneck factor by $bf$. We can reduce parameters if we use two-layered TDFs of $bf > 2$. We investigate the effect of adding additional layers in \S \ref{sec:exptimedistributed}.

\subsubsection{Time-Distributed Convolutions}

We propose an alternative time-distributed block named Time-Distributed Convolutions (TDC), which is applied separately and identically to each multi-channeled frame. 
It is a series of 1-D convolution layers.
Inspired by \cite{mmdensenet,mmdenselstm}, it takes form of a \textit{dense block} \cite{densenet} structure. A dense block consists of densely connected composite layers, where each composite layer is defined as three consecutive operations: 1-D convolution, BN, and ReLU.
As discussed in \cite{densenet, mmdensenet, mmdenselstm} the densely connected structure enables each layer to propagate the gradient directly to all preceding layers, making a deep CNN training more efficient.

\begin{figure}[htbp]

    \centering
    \includegraphics[width=\columnwidth]{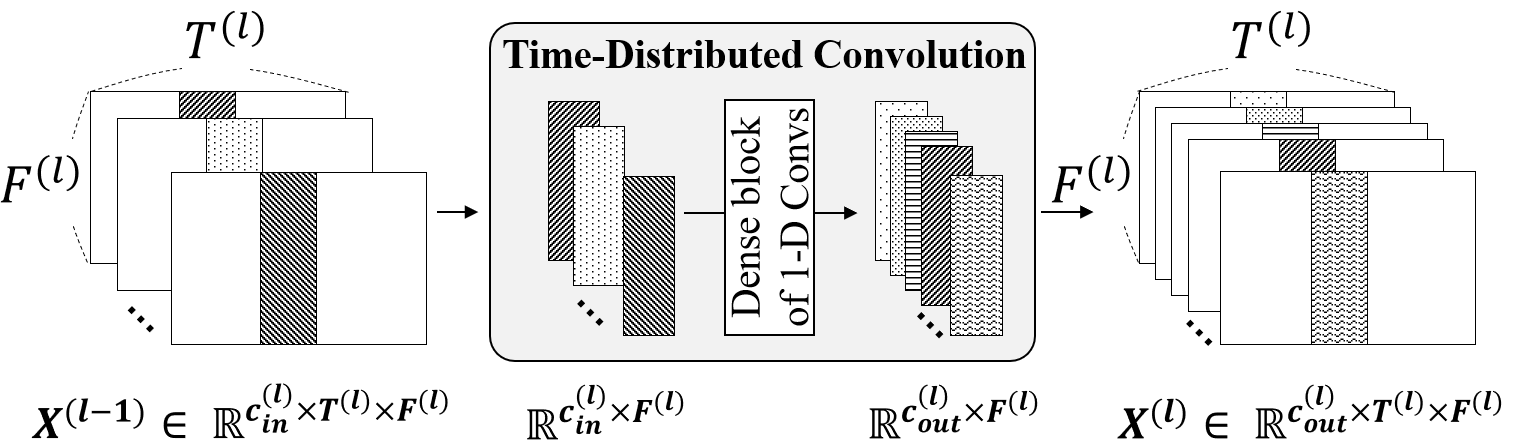}

    \caption{Time-Distributed Convolutions}
    \label{fig:tdc}
\end{figure}

\subsection{Time-Frequency Blocks}

The performances of U-Nets with time-distributed blocks were above our expectation (see \S \ref{sec:exptimedistributed}), but were still inferior considerably to those of current SOTA methods. The reason is that features observed in musical sources include sequential patterns (e.g., vibrato, tremolo, and crescendo) or musical patterns (e.g., rhythm, repetitive structure), which cannot be modeled by time-distributed blocks.

While time-distributed blocks cannot model the temporal context, time-frequency blocks try to extract features considering both the time and the frequency dimensions.
We introduce the Time-Frequency Convolutions (TFC) block, which is used in \cite{mmdensenet}. 
We also propose two novel blocks that combine two different transformations.

\subsubsection{Time-Frequency Convolutions}

\begin{figure}[htbp]
    \centering
    \includegraphics[width=\columnwidth]{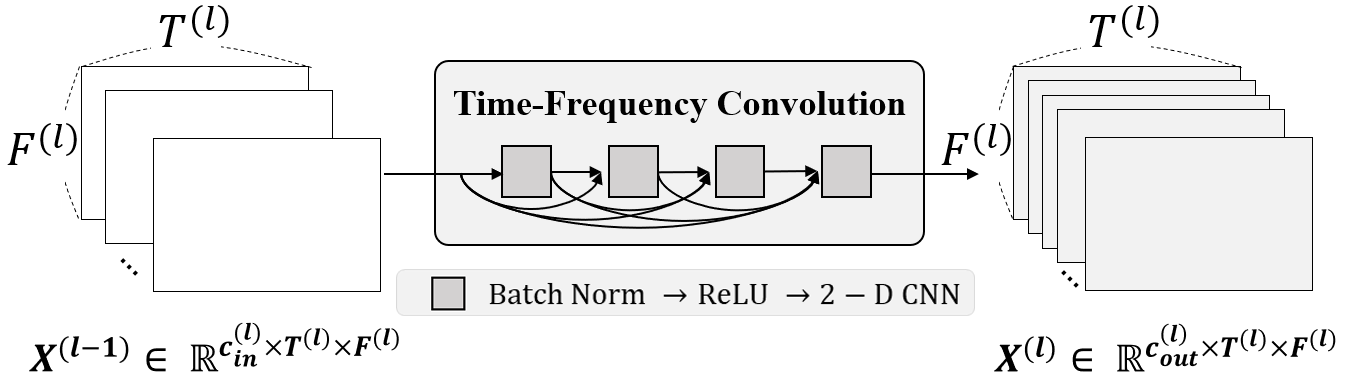}
    \caption{Time-Frequency Convolutions}
    \label{fig:tfc}
\end{figure}

The Time-Frequency Convolutions (TFC) is a dense block of 2-D CNNs, as shown in Figure \ref{fig:tfc}.
The dense block consists of densely connected composite layers, where each layer is defined as three consecutive operations: 2-D convolution, BN, and ReLU.   
It is applied to the spectrogram-like input representation in the time-frequency domain.
Every convolution layer in a dense block has kernels of size $(k_F, k_T)$. Its 2-D filters are trained to jointly capture features along both frequency and temporal axes.

\subsubsection{Time-Frequency Convolutions with TDF}
\label{sec:TFCTDF}

\begin{figure}[htbp]

\centering
\includegraphics[width=\columnwidth]{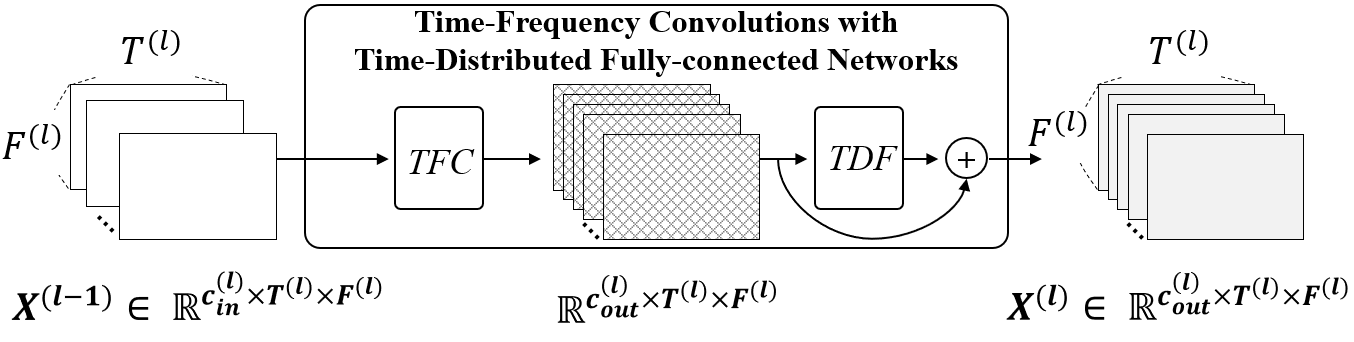}

\caption{Time-Frequency Convolutions with TDF}
\label{fig:tfctdf}

\end{figure}

We propose the Time-Frequency Convolutions with Time-Distributed Fully-connected networks (TFC-TDF) block. 
It utilizes two different blocks inside: a TFC block and a TDF block.
Figure \ref{fig:tfctdf} describes a TFC-TDF block.
It first maps the input $X^{(l-1)}$ to a same-sized representation with $c_{out}^{(l)}$ channels by applying the TFC block. Then the TDF block is applied to the dense block output. A residual connection is also added for efficient gradient flow.


Phasen \cite{phasen} has shown that inserting time-distributed operations into intermediate blocks can improve speech enhancement performance.
We validate whether it also works for SVS or not in \S \ref{sec:exptimevariant}.

\subsubsection{Time-Distributed Convolutions with RNNs}
\label{sec:tdcrnn}

We propose an alternative way to consider both the time and frequency dimensions.
A Time-Distributed Convolutions with Recurrent Neural Networks (TDC-RNN) block uses two different blocks: a TDC block for extracting timbre features and RNNs for capturing temporal patterns.
It extracts timbre features and temporal features \textit{separately}, unlike a TFC block. We validate whether this approach can outperform the 2-D CNN approach by comparing TDC-RNNs with TFCs in \S \ref{sec:exptimevariant}.

The structure of a TDC-RNN block is similar to that of a TFC-TDF block.
It applies the TDC block to an input $X^{(l-1)}$, and obtains a same sized hidden representation with  $c_{out}^{(l)}$ channels. The RNNs compute the hidden representation and output an equally sized tensor. A residual connection is added, as is a TFC-TDF block.



\section{Experiment}
\label{sec:exp}

We evaluate U-Nets with different types of blocks introduced in \S \ref{sec:blocks}. 
We compare the performance of models in \S \ref{sec:exptimedistributed} and \S \ref{sec:exptimevariant}. Also, we compare our models with SOTA models in \S \ref{sec:sota}.
We compare the spectrogram estimation framework in \S \ref{sec:cvsm}.
We discuss reusable insights in \S \ref{sec:discussion}.

\subsection{Setup}

\subsubsection{Dataset}

Train and test data were obtained from the MUSDB dataset \cite{musdb18}.
The train and test sets of MUSDB have 100 and 50 musical tracks each, all stereo and sampled at 44100 Hz. 
Each track file consists of the mixture and its four source audios: `vocals,' `drums,' `bass' and `other.' 
Since we are evaluating on singing voice separation, we only use the `vocals' source audio as the separation target for each mixture track. 

\subsubsection{Model Configurations}

We implemented U-Nets with different blocks (\S \ref{sec:blocks}).
Each model is based on the U-Net architecture (\S \ref{sec:unet}) on the CaC framework (\S \ref{sec:cac}).
We set $c_{in}^{(1)}$, the number of internal channels to be 24, as mentioned in \S \ref{sec:unet}.
Each model uses a single type of block for its intermediate blocks.
We usually used an FFT window size of 2048 and a hop size of 1024 for STFT. However, we used a larger window size in some models for a fair comparison with SOTA methods.

\subsubsection{Training and Evaluation}

Weights of each model were optimized with RMSprop \cite{rmsprop} with learning rate  $lr \in [0.0005, 0.001]$ depending on model depth. 
Each model is trained to minimize the mean square error between $\hat{T}$ and $\mathbf{T}$ as mentioned in \S \ref{sec:cac}.
We use the default validation set (14 tracks) as defined in the \textit{MUSDB} package, and use the Mean Squared Error (MSE)  between target and estimated signal (waveform) as the validation metric for validation. Data augmentation \cite{blend} was done on the fly to obtain fixed-length mixture audio clips comprised of the source audio clips from different tracks.

We use the official evaluation tool\footnote{https://github.com/sigsep/sigsep-mus-eval} provided by the organizers of the SiSEC2018\cite{sisec} to measure Source-to-Distortion Ratio (SDR) \cite{bss}.
We use the median SDR value over all the test set tracks to obtain the overall SDR performance for each run, as done in the SiSEC2018.
We report the average of `median SDR values' over three runs for each model.

\subsection{U-Nets with Time-Distributed Blocks}
\label{sec:exptimedistributed}

We implemented and trained U-nets with TDC and TDF blocks. 
We also implemented models with TDC blocks that do not use down/up-sampling to investigate the effect of down/up-sampling in the frequency axis.
The other models use 1-D convolution/transposed-convolution layers with stride 2 for down/up-sampling.
Every TDC block is a dense block with 5 composite layers with the growth rate 24 (used in dense blocks \cite{densenet}). The kernel size of each convolution layer in a dense block is 3.
Each TDF block is either single-layered or two-layered.
The bottleneck factor $bf$ of each TDF block is set to be 4.
All models have 17 intermediate blocks except for two shallow models.

\begin{table}[t]
\begin{center}
\begin{tabular}{|l|c|r|c|}
    
    \hline 
    block type  & \# {blocks} & \# {params} & {SDR} \\ \hline 
    TDC (w/ sampling) & 17  & 0.54M & \textbf{4.86} \\ \hline
    TDC (w/o sampling) & 17 & 0.52M     & 3.78 \\ \hline
    TDC (w/o sampling) & 3 & 0.09M     & 3.56 \\ \hline
    
    TDF (w/o hidden layer) & 17 & 2.83M & 4.75 \\ \hline
    TDF (w/ hidden layer) & 17 & 1.44M & 4.05 \\ \hline
    TDF (w/ hidden layer) & 3 & 1.19M & 4.01 \\ \hline
    
    \end{tabular}
    
    \caption
        {Evaluation results of Time-Distributed Blocks.
        \label{table:timedistributed}
        }
\end{center}
\end{table}

We summarize evaluation results in Table \ref{table:timedistributed}. The TDC block-based U-Net with sampling achieves an SDR of 4.86, the highest among the three models.
Results show that the use of down/up-sampling in TDC-based U-Nets was significant, although the model without sampling can exploit higher resolution of internal representations.
It may indicate that enlarging receptive fields via sampling may help the model to capture long-term dependencies better, and long-term dependencies are preferred over local features when distinguishing unique time-independent frequency patterns.
(at least for these configurations). 

Although FCNs can capture long-ranged patterns along the frequency domain, as mentioned in \cite{phasen}, TDF-based U-Nets did not perform well enough compared to the TDC-based models in a deep architecture.
Among TDF-based models, the U-Net equipping single-layered TDFs (the fourth row of Table \ref{table:timedistributed}) outperforms the other models.
However, it is notable that we can reduce parameters when we use two-layered TDFs.
Also, we found that the TDF blocks can outperform  TDC blocks in a shallow architecture (the third and sixth row of Table \ref{table:timedistributed}). The reason is that the U-Nets with few TDC blocks has a small receptive field, while a single TDF block has a full receptive field in the frequency dimension, which has led us to inject it in a time-frequency block instead of TDC (see \S \ref{sec:TFCTDF}).


\subsection{U-Nets with Time-Frequency Blocks}
\label{sec:exptimevariant}

We implemented U-Nets with time-frequency blocks. 
All models are trained on 3 seconds (128 STFT frames) of music. Since the number of frequency bins is much larger than the number of frames, models with more than 7 neural transforms use both $2 \times 2$ or $2 \times 1$ sized down/up-sampling layers to scale the frequency axis more than 3 times while maintaining the number of scales in the temporal axis to 3. Exceptionally, we use different down/up-sampling layers for one model to investigate the effect of down/up-sampling in the temporal axis.

\begin{table}[t]
\begin{center}

\begin{tabular}{|l|c|c|r|c|}
\hline

model & sampling & {\# blocks} &  {\# params} & SDR \\ \hline
    
TFC & O & 17 & 1.56M & 6.89
\\ \hline

TFC & X &  17 & 1.56M & 6.75 \\ \hline
TDC-RNN & O & 17 & 2.08M & 6.69 \\ \hline 
TFC-TDF & O & 7 & 0.99M  & 7.07 \\ \hline 
TFC-TDF & O & 17 & 1.93M  & \textbf{7.12}   \\ \hline
\end{tabular}

{\caption{Evaluation results of Time-Frequency Blocks.}
 
\label{table:timevariant}}

\end{center}
\end{table}

We set every TFC block to have 5 convolution layers with kernel size $3 \times 3$. We set the growth rate to be 24,  the same growth rate of \S \ref{sec:exptimedistributed}. 
By using this TFC block configuration, we implemented a TFC-based U-Net (the first row of Table \ref{table:timevariant}).
We set the model in the second row to use different down/up-sampling layers to investigate the effect of down/up-sampling in the temporal axis.
Every kernel size used in each down/up-sampling layer of this model is $2 \times 1$ to preserve the temporal resolution while scaling frequency resolution. 
The first two rows of Table \ref{table:timevariant} summarize the experiment results of two TFC-based models.
The model that preserves the temporal resolution was slightly inferior to the other model.
It is also notable that our U-Nets with TFC blocks achieve comparable results with state-of-the-art methods \ref{sec:sota}, even using lower frequency resolution. 
Compared to the frequency axis where the TDC-based U-Net with down/up-sampling outperforms the counterpart model, no significant SDR was gained by enlarging the receptive field by down/up-sampling.

We reused the configuration of TDC in of \S \ref{sec:exptimedistributed}, for TDCs in TDC-RNN blocks.
The RNN layers were implemented with bidirectional GRUs with a single hidden layer, which has $f/16$ hidden units, where $f$ is the number of input frequency bins. Although having more parameters and a better potential for capturing long temporal dependencies than the two fully convolutional models, TDC-RNN performs lower than them.
Increasing the number of hidden units or hidden layers could have increased SDR since many other state-of-the-art recurrent models use a hidden size that is at least 512.
Increasing the number of STFT frames, thus training on longer clips of music, might have also worked.
Although it performs the worst among the time-frequency blocks, it is superior to all the time-distributed blocks. It indicates that inter-frame operations are necessary for higher quality separation.

The fourth and fifth rows of the Table \ref{table:timevariant} shows promising results regarding the U-Nets with TFC-TDF blocks. 
We reused the same TFC setting above, and we set $bf$ to be 16 for each TDF.
The 7-blocked U-Net with TFC-TDFs outperforms the other 17-blocked models. 
These results show that inserting FCNs into intermediate blocks can be useful for MSS as well as for Speech Enhancement \cite{phasen}. Also, results show that it is also achievable with fewer parameters by using FCNs with a bottleneck layer.

\subsection{Comparison with SOTA models}
\label{sec:sota}

We compare our models with other spectrogram-based models on the MUSDB benchmark. 
The first three rows of Table \ref{table:eval} shows the SDR performance of SOTA models, namely DGRU-DGConv \cite{dilatedlstm}, TAK1 \cite{mmdenselstm}, and UMX \cite{UMX}. Their SDRs can be found in \cite{dilatedlstm}, SiSEC2018 repository\footnote{https://github.com/sigsep/sigsep-mus-2018}, and UMX repository\footnote{https://github.com/sigsep/open-unmix-pytorch}.
We estimated the lower bound of the number of parameters of DGRU-DGConv with 1-D CNN parameters without considering its GRUs. 

Comparing with Table \ref{table:timevariant}, we can see that our models perform comparably to or even outperform existing models even with less frequency resolution and fewer parameters. 
On top of that, our TFC extensions do not use recurrent layers, which is a key factor in the other previous models. It may lead to shorter forward/backward propagation time.
Also, it is worth noting that previous models adopt Multi-channel Wiener Filtering as a post-processing method to further enhance SDR. Ours directly use the signal reconstruction output without such post-processing.

For a fair comparison with SOTA models, we trained an additional U-Net with 9 TFC-TDF blocks (notated as `large' in Table \ref{table:eval}) with the same frequency resolution as the other SOTA models (FFT window size = 4096) and achieved outstanding results with a 0.9 dB gain over DGRU-DGConv. 

\begin{table}
    \begin{center}
\begin{tabular}{|c|c|c|}

\hline
\multicolumn{1}{|c|}{model} & \# parameters & SDR (vocals) \\ \hline
DGRU-DGConv               & more than 1.9M           & 6.99                    \\  \hline
TAK1                      & 1.22M           & 6.60                    \\  \hline 
UMX  & 8.89M           & 6.32                    \\  \hline \hline
TFC-TDF (small) & 0.99M & \textbf{7.07} $\pm .08$
\\  \hline
TFC-TDF (large) & 2.24M & \textbf{7.98} $\pm .07$
\\ \hline

\end{tabular}
        {\caption{Comparison: SDR median value on test set. 
            \label{table:eval}}
        }
    \end{center}
\end{table}

\subsection{Spectrogram Estimation: Complex vs Magnitude}
\label{sec:cvsm}

\begin{table}
\begin{center}

\begin{tabular}{|c|c|c|c|c|}

\hline 
esimation   & {n\_fft} & \# {blocks}  & \# {params} & {SDR} \\ \hline
CaC & 2048 & 7 & 0.99M & \textbf{7.07} \\ \hline
Mag & 2048 & 7 & 0.99M & 6.43 \\ \hline 
CaC & 4096 & 9 & 2.24M & \textbf{7.98} \\ \hline
Mag & 4096 & 9 & 2.24M & 7.24 \\ \hline

\end{tabular}
{\caption{Comparison of TFC-TDFs: CaC vs Mag
\label{table:cvsm}}}
\end{center}
\end{table}

For our final experiment, we see how much SDR was gained by extending a magnitude-only model into a CaC model. Our TFC-TDF-based U-Nets in Table \ref{table:cvsm} are compared to their magnitude-only form (referred to as `Mag'). They use the same hyperparameter set except for $c_{in}^{(0)}$, the input/output number of channels. Mag also has an additional ReLU after the final $1\times 1$ convolution to obtain non-negative-valued output spectrograms. Results show that training with raw STFT outputs instead of magnitudes significantly boosts SDR performance.
It is also notable that the Mag model with n\_fft of 4096 still outperforms all previous state-of-the-art models in Table \ref{table:eval}. 

\subsection{Discussion: Developing Reusable Insights}
\label{sec:discussion}
Our work provides a practical guideline for choosing  fundamental building blocks to develop an SVS or MSS model based on the U-Net architecture as follows.
\begin{itemize}
    \item TDC-based models are sensitive to the number of blocks, compared to TDF-based models.
    \item Using down/up-sampling is important for CNN-based blocks, especially in the frequency dimension. 
    \item Stacking 2-D CNNs is a simple but effective way to capture T and F features, compared to TDC-RNNs. 
    \item Injecting a time-distributed block to a time-frequency block can improve SDR. 
    \item A simple extension from a magnitude-only U-Net to a CaC U-Net can improve SDR. 
\end{itemize}

Our work is not limited to the U-Net-architecture nor MSS. 
Blocks can be used as core components in more complex architectures as well. We can use different types of blocks for a single model, meaning that a lot of space remains for improvement.
Also, our observations can be exploited in other MIR tasks such as Automatic Music Transcription (AMT) or Music Generation: for example, we expect that injecting TDFs to intermediate blocks for $f_0$ estimation model can improve performance since fully-connected layer can efficiently model long-range correlations such as harmonics.

\section{Conclusion and Future works}

In this paper, we designed several types of blocks based on different design strategies.
We implemented U-Net models with these blocks for SVS and evaluated their performance.
Our experiments provide abundant material for future works by comparing several U-Nets with different types of blocks.
Also, one of our models outperforms SOTA methods.
For future work, we would like to extend this model to utilize attention networks for modeling long-term dependencies observed in both the frequency and the temporal axis.

\section{ACKNOWLEDGEMENTS}
This work was supported by the National Research Foundation
of Korea(NRF) grant funded by the Korea government(MSIT) (No.
NRF-2019R1F1A1062719, NRF-2020R1A2C1012624).

\bibliography{svs}

%
%
%
%

\end{document}